\documentclass[twocolumn,aps,floats,superscriptaddress]{revtex4}
\usepackage{amsmath}
\usepackage{graphicx}

\newcommand{\mr}[1]{{{\mathrm{#1}}}}
\newcommand{\mcal}[1]{{\mathcal{#1}}}
\newcommand{\dsd}{d_{\sigma}^{\dagger}}
\newcommand{\ds}{d_{\sigma}}
\newcommand{\fsd}{f_{\sigma}^{\dagger}}
\newcommand{\fs}{f_{\sigma}}
\newcommand{\disd}{d_{\mathrm{i}\sigma}^{\dagger}}

\newcommand{\fisd}{f_{\mathrm{i}\sigma}^{\dagger}}
\newcommand{\fis}{f_{\mathrm{i}\sigma}}

\newcommand{\djs}{d_{\mathrm{j}\sigma}}
\newcommand{\fjsd}{f_{\mathrm{j}\sigma}^{\dagger}}
\newcommand{\fjs}{f_{\mathrm{j}\sigma}}
\newcommand{\s}{\sigma}

\newcommand{\w}{\omega}
\newcommand{\vk}{{\bf k}}

\newcommand{\under}[1]{_{|_{\scriptstyle#1}}}
\newcommand{\dt}{\partial_\tau}
\newcommand{\eps}{\epsilon_{0}}
\newcommand{\ek}{\epsilon_{\bf{k}}}
\newcommand{\inte}{\int_0^\beta \!\!\!\! \mr{d}\tau}
\newcommand{\inteps}{\int_{-D}^{D} \!\! \mr{d} \epsilon \;}
\def\ket#1{\left\vert #1 \right\rangle}

\def\qbar{\overline{Q}}

\def\hQ{\hat{Q}}
\def\hL{\hat{L}}
\def\veff{V_{\mr{eff}}}
\def\geff{\Gamma_{\mr{eff}}}

\def\hop{t_{\mr{ij}}}
\def\hopeff{t^{\mr{\,eff}}_{\,\mr{ij}}}
\def\Jeff{{\cal J}^{\mr{\,eff}}_{\,\mr{ij}}}
\def\avcos{\langle\cos\theta\rangle}
\def\iomn{i\omega_n}
\def\epsbar{\overline{\epsilon}\,}
\def\gap{\Delta_g}
\def\abseps{|\epsbar|}
\def\ucinf{U_{c}^{\infty}}
\def\deltainf{\Delta}
\def\qbar{Q_X}

\begin{document}

\title{Slave-rotor mean field theories of strongly correlated systems\\
and the Mott transition in finite dimensions}
\author{Serge Florens}
\affiliation{Institut f\"ur Theorie der Kondensierten Materie,
Universit\"at Karlsruhe, 76128 Karlsruhe, Germany}
\affiliation{Laboratoire de Physique Th\'eorique, Ecole Normale
Sup\'erieure, 24 rue Lhomond, 75231 Paris Cedex 05, France}
\author{Antoine Georges}
\affiliation{Centre de Physique Th\'eorique, Ecole Polytechnique, 91128 Palaiseau Cedex}
\affiliation{Laboratoire de Physique Th\'eorique, Ecole Normale
Sup\'erieure, 24 rue Lhomond, 75231 Paris Cedex 05, France}

\begin{abstract}
\vspace{0.3cm}
The multiorbital Hubbard model is expressed in terms of quantum phase variables
(``slave rotors'') conjugate to the local charge, and of auxiliary fermions, providing an
economical representation of the Hilbert space of strongly correlated systems.
When the phase variables are treated in a local mean-field manner,
similar results to the dynamical mean-field theory are obtained, namely a
Brinkman-Rice transition at commensurate fillings together with a ``preformed'' Mott gap
in the single-particle density of states.
The slave- rotor formalism allows to go beyond the local description and take
into account spatial correlations, following an analogy to the superfluid-insulator
transition of bosonic systems.
We find that the divergence of the effective mass at the metal-
insulator transition is suppressed by short range magnetic correlations in finite-
dimensional systems.
Furthermore, the strict separation of energy scales between the Fermi-
liquid coherence scale and the Mott gap, found in the local
picture, holds only approximately in finite dimensions, due to
the existence of low-energy collective modes related to zero-sound.
\end{abstract}

\maketitle


\section{Introduction}

Strongly correlated fermion systems constitute a challenge, both from a
fundamental point of view (with phenomena such as the Mott transition~\cite{RMP_MIT}
and high- temperature superconductivity), and on a more quantitative level
with the need of reliable tools to handle intermediate and strong coupling regimes
(even for simplified models such as the Hubbard model).
In recent years, the dynamical mean-field theory (DMFT) has allowed for
significant progress in this respect~\cite{RMP_DMFT}.
In particular, this approach has led to a detailed theory of the Mott transition, and to
a quantitative description of the physics of strongly correlated metals.
Despite these successes, the limitations of this approach have been emphasized
on many occasions. The main one has to do with the effect of spatial correlations
(e.g magnetic short-range correlations), and
more precisely with the effect of these correlations on the properties of
quasiparticles. For example, the tendency to form singlet bonds due to superexchange is
widely believed to be a key physical effect in weakly doped Mott insulators.
Also, at the technical level, the application of DMFT to materials with
a large orbital degeneracy (e.g in combination
with ab-initio methods~\cite{held_review_lda+dmft_1,lichtenstein_magnetism_2002,georges_strong}),
as well as cluster extensions of DMFT~\cite{RMP_DMFT,maier_cluster_review} are
computationally challenging because they involve
the solution of a multi-orbital quantum impurity model.

For these reasons, there is still a strong need for approximate, simpler
treatments of strongly correlated fermion models. Those treatments should
incorporate some of the DMFT successes (e.g regarding the description
of the metal-insulator transition), but they should also pave the road for
describing physical effects beyond DMFT at least at a qualitative level.

The purpose of this paper is to present a simple mean field description of
correlated systems which fullfills some of these goals. Our main idea
is to focus on the degrees of freedom associated to the relevant physical
variable associated to the Mott transition, namely a slave quantum rotor field,
dual to the local electronic charge. This slave rotor representation was
introduced previously by us for the description of quantum impurity
models and mesoscopic devices~\cite{SF_AG1,florens_qdot_prb},
and is applied here in the context of lattice models.
This allows for a simple reformulation
of the orbitally degenerate Hubbard model, which,
when the interaction has full orbital
symmetry, is quite superior to previously developed slave- boson
representations~\cite{Kot_Ruck,Hasegawa,florens_orbital_2002_prb}.

When the simplest possible (single-site) mean-field approximation is used
in conjunction with this slave-rotor representation, a description of the
Mott transition very similar to that of DMFT is found.
The metallic phase disappears through a Brinkman-Rice transition, at
which the quasiparticle wieight vanishes and the effective mass diverges.
The slave rotor approach does preserve Hubbard bands in the insulator,
and a ``preformed'' Mott spectral gap opening up discontinuously at the
transition is found, as in DMFT.

The most interesting aspect of our approach lies however in the possibility of
going beyond this purely local mean-field description.
By decoupling the spinons and slave rotor degrees of freedom, the Hubbard model
is mapped onto a free spinon hamiltonian self-consistently coupled to a
quantum XY lattice model. The (dis)ordering transition of the latter corresponds
to the Mott transition, in analogy with the superfluid-insulator transition
of the bosonic Hubbard model. Because spatial correlations are now included,
we find important
modifications to the DMFT picture. In particular, the effective mass
remains finite at the transition, due to the quenching
of the macroscopic entropy by magnetic correlations in the Mott phase.
Importantly, low-energy charge collective modes are shown to affect 
the opening of the Mott gap, which now develops in a continuous manner, 
so that the separation of energy scales found in DMFT only holds in an approximate manner.
These simple results can be considered as deviations from the DMFT predictions
that could possibly be observed in photoemission experiments~\cite{perfetti}.
However, restauration of the local gauge symmetry should occur due
to fluctuations beyond the mean-field approximation, possibly modifying the
latter result on a qualitative way. 

The paper is organised as follows. In section~\ref{sec:repr} we introduce the
exact slave rotor description of a simple atomic level with orbital degeneracy,
and show that an
approximate treatment of the local constraint is sufficient to describe
correctly the full Coulomb staircase, as well as one-particle spectra.
Then, in section~\ref{sec:simplest}, we develop the simplest (local) mean field
treatment of both the Anderson and Hubbard models, and in the latter case, study
the multiorbital Mott transition. Finally, spatial fluctuations beyond DMFT
are included in section~\ref{sec:spatial}, with an emphasis on the behavior
of the effective mass and the excitation spectrum. The conclusion presents
several possible applications and extensions of our formalism, and also discusses
some of the open issues raised by our results.

\section{Rotor representation of interacting fermions}
\label{sec:repr}

\subsection{Slave-rotor representation}

In Ref.~\cite{SF_AG1} (see also~\cite{florens_phd}),
we introduced a representation of the
Hilbert space of $N$ fermions $\dsd$ in terms of a collective
phase degree of freedom $\theta$, conjugate to the total charge,
and of $N$ auxiliary fermions $\fsd$. The spin/orbital index runs
over $N$ values $\sigma=1 \ldots N$ (e.g $\s=\uparrow,\downarrow$
for $N=2$). In the following, we consider only interactions which
have the full $SU(N)$ spin/orbital symmetry. Let us consider the
Hamiltonian corresponding to a single ``atomic level'', in the
presence of a local Hubbard repulsion:
\begin{equation}
\label{eq:Hat}
H_{\mr{at}} = \sum_\s \eps\, \dsd \ds + \frac{U}{2} \left[ \sum_\s \dsd \ds
  -\frac{N}{2} \right]^2
\end{equation}
The crucial point is that the spectrum of the atomic Hamiltonian
(\ref{eq:Hat}) depends only on the total fermionic charge $Q=0,\cdots,N$ and has
a simple quadratic dependence on $Q$:
\begin{equation}
\label{eq:atomic_spectrum}
E_Q = \eps Q + \frac{U}{2} \left[ Q -\frac{N}{2} \right]^2
\end{equation}
There are $2^N$ states, but only $N+1$ different energy levels, with degeneracies
$\binom{N}{Q}$.
In conventional slave boson methods~\cite{Kot_Ruck,florens_orbital_2002_prb},
a bosonic field is introduced for {\it each} atomic state $|\s_1\cdots \s_Q\rangle$ (along with
spin-carrying auxiliary fermions $\fsd$). Hence, these methods are not describing
the atomic spectrum in a very economical manner, and lead to very tedious calculations
when orbital degeneracy becomes large, even at the mean-field level
\cite{Hasegawa,Fresard_Kotliar,Bunemann,florens_orbital_2002_prb}. However, we
stress that the extensive number of degrees of freedom necessary in those other
approaches can become useful when the $SU(N)$ symmetry is broken,
either by magnetic order or crystal fields.

The spectrum of (\ref{eq:Hat}) can actually be reproduced by introducing,
besides the set of auxiliary fermions $\fsd$,
a {\it single} additional variable, namely the angular momentum
$\hat{L} = -i \partial/\partial\theta$ associated with a quantum $O(2)$ rotor
$\theta$, an angular variable in $[0,2\pi]$. Indeed, the energy levels
(\ref{eq:atomic_spectrum}) can be obtained using the following Hamiltonian
\begin{equation}
\label{eq:Hloc_f_theta}
H_{\mr{at}} = \sum_\s \eps \fsd \fs + \frac{U}{2} \hat{L}^2
\end{equation}
A constraint must be imposed, which insures that the total number of fermions
is equal to the $O(2)$ angular momentum (up to a shift, in our conventions):
\begin{equation}
\label{eq:contrainte}
\hat{L} = \sum_\s \left[\fsd \fs - \frac{1}{2} \right]
\end{equation}
This restricts the allowed values of the angular momentum to be
$\ell=Q-N/2=-N/2,-N/2+1,\cdots,N/2-1,N/2$, while in the absence of any
constraint $\ell$ can be an arbitrary (positive or negative) integer.
The spectrum of (\ref{eq:Hloc_f_theta}) is $\eps Q + U\ell^2/2$, with $\ell=Q-N/2$
thanks to (\ref{eq:contrainte}), so that it coincides with (\ref{eq:atomic_spectrum}).

It is easily checked that the full Hilbert space is correctly described as:
\begin{equation}
\ket{\s_1\ldots \s_Q}_d \,=\,
\ket{\s_1\ldots \s_Q}_f \ket{\ell=Q-N/2}_\theta
\end{equation}
in which $\ket{\s_1\ldots \s_Q}_{d,f}$ denotes the antisymmetric fermion state
built out of $d$- and $f$-fermions, respectively, and $\ket{\ell}_\theta$
denotes the quantum rotor eigenstate with angular momentum $\ell$,
{\it i.e.} $\langle\theta \ket{\ell}_\theta = e^{i \ell \theta}$.
For $N=2$, this corresponds to:
$\ket{\uparrow}_d = \ket{\uparrow}_f \ket{0}_\theta$,
$\ket{\downarrow}_d = \ket{\downarrow}_f \ket{0}_\theta$,
$\ket{\uparrow\downarrow}_d = \ket{\uparrow\downarrow}_f \ket{+1}_\theta$
and $\ket{0}_d = \ket{0}_f \ket{-1}_\theta$.
The creation of a physical electron with spin $\sigma$ is associated to the
action of $\fsd$ on such a state as well as raising the total charge
(angular momentum) by one unit. Since the raising operator
is $e^{i\theta}$, this leads to the representation:
\begin{equation}
\label{eq:repres}
\dsd \equiv \fsd\, e^{i \theta}\,\,\,,\,\,\,
\ds \equiv \fs\, e^{-i\theta}
\end{equation}
The key advantage of the quantum rotor representation is that the original
quartic interaction between fermions has been replaced in (\ref{eq:Hloc_f_theta})
by a simple kinetic term for the phase field, $(U/2)\hat{L}^2$.

We point out here that a similar phase representation was developed before in
the context of Coulomb blockade in mesoscopic systems, see e.g~\cite{grabert,schoeller,lebanon}.
However, the present work and our previous paper~\cite{SF_AG1} present the first
applications of the rotor technique to the context of strongly correlated lattice
models.
In particular, the question of quasiparticle coherence which is crucial to the description of
a Fermi liquid cannot be investigated seriously with a phase-only description
\cite{SET_QMC}, as shown in~\cite{florens_qdot_prb}. In this perspective, the
slave rotor should be seen as a natural extension (and simplification) of the usual slave boson
techniques~\cite{slave_boson1,kotliar_houches} in the context of a finite but orbitally
symmetric Coulomb repulsion. In principle, it can also be applied to systems with
long-range interactions~\cite{SF_LdM_AG,florens_phd}.

\subsection{Treating the constraint on average: atomic limit}

In the following, we will study different kinds of mean-field approximations
based on this slave-rotor representation. A common trait of these mean-field
approximations is that the number constraint (\ref{eq:contrainte}) will be treated
{\it on average}. This is equivalent to treating the constraint in a ``grand-canonical''
ensemble, which would of course be exact in the limit of a large
spin/orbital degeneracy $N\rightarrow\infty$. In this section, we investigate
the accuracy of this approximation for the atomic Hamiltonian (\ref{eq:Hat}),
for finite values of $N$.

\subsubsection{Coulomb staircase: occupancy vs. $\eps$}

\begin{figure}[htbp]
\begin{center}
\includegraphics[width=8cm]{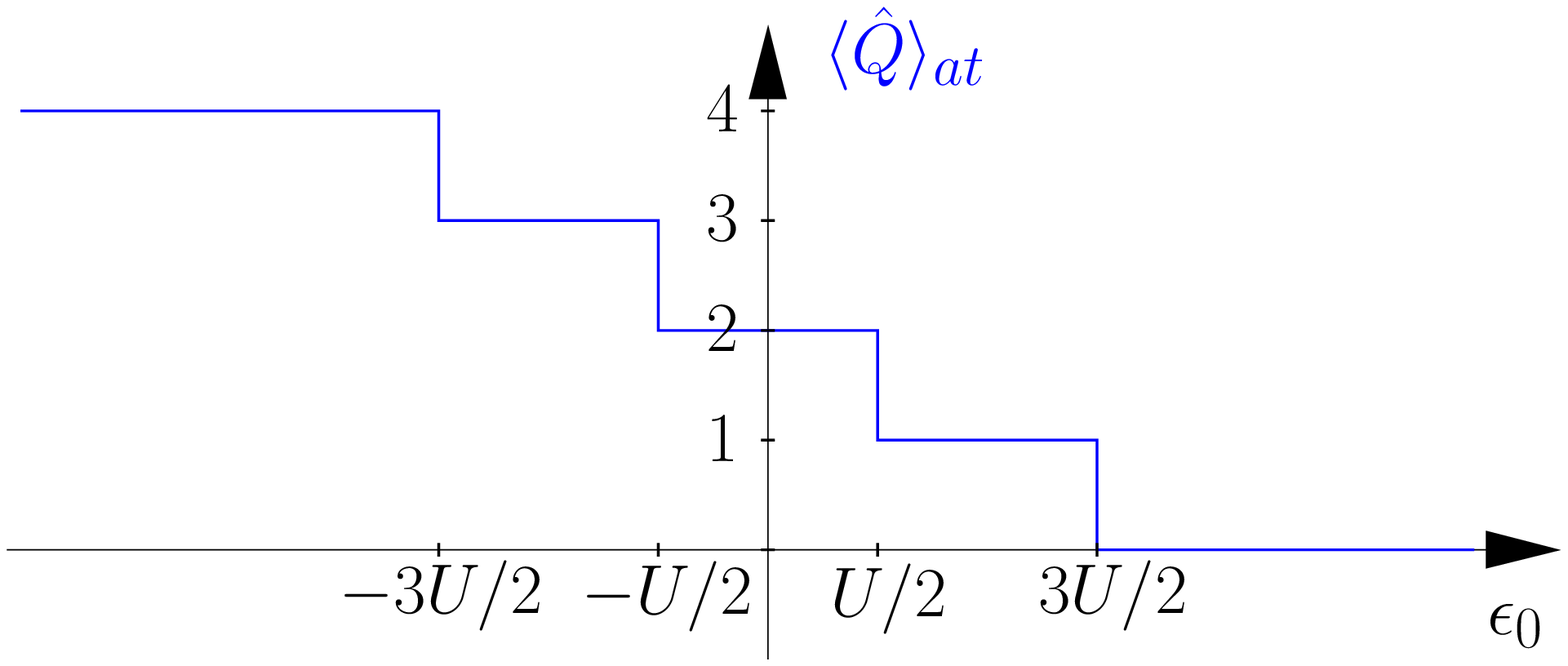}
\end{center}
\caption{Coulomb staircase in the atomic limit for the case of two orbitals, $N=4$.}
\label{fig:staircase}
\end{figure}

\begin{figure}[htbp]
\begin{center}
\includegraphics[width=7.5cm]{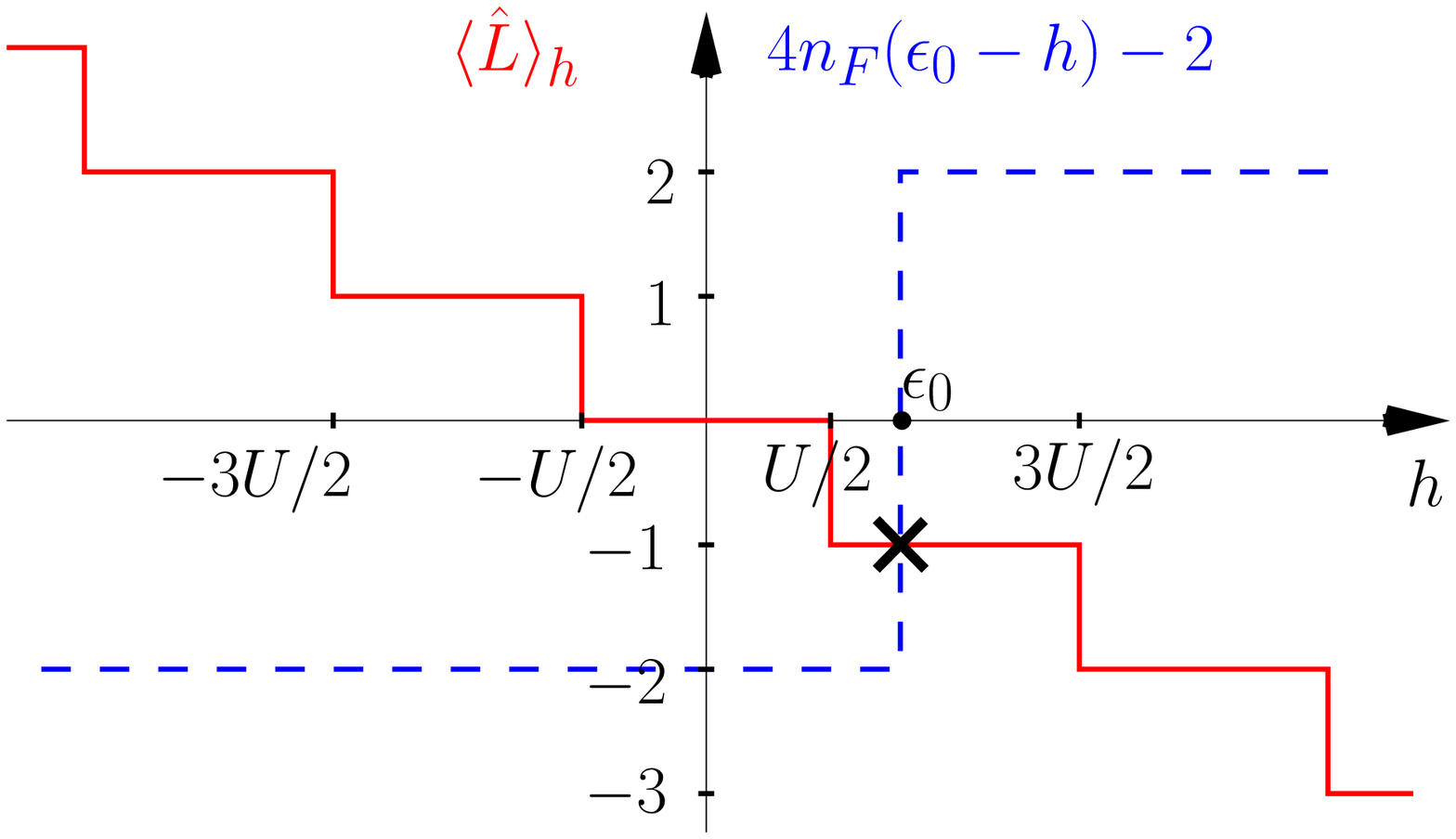}
\end{center}
\caption{Graphical solution of the average constraint
equation~(\ref{eq:atomic_h}). The intersect (cross) moves exactly along the Coulomb
staircase shown in figure~\ref{fig:staircase}.}
\label{fig:solve_constraint}
\end{figure}

Let us first consider the dependence of
the average occupancy
$\langle\hQ\rangle=\langle\sum_\s \dsd\ds\rangle$
on the position of the atomic level $\eps$, which reads:
\begin{equation}
\langle \hQ \rangle_{at} = \frac{1}{Z_{at}}\,
\sum_{Q=0}^{N} \binom{N}{Q}\,Q\,e^{-\beta\,E_Q}
\label{eq:exact_charge}
\end{equation}
with $Z_{at}=\sum_Q \binom{N}{Q} e^{-\beta E_Q}$. In the limit of zero temperature, the dependence
of $\langle \hQ \rangle$ on $\eps$ is the ``Coulomb staircase'' in Fig.~\ref{fig:staircase}.
When treating the constraint on average, a Lagrange multiplier $h$ is introduced which
is conjugate to (\ref{eq:contrainte}), and one optimizes over $h$ instead of
fully integrating over it. This amounts to consider the following effective
Hamiltonians:
\begin{eqnarray}
H_f^{at} & = & (\eps-h)\,\sum_{\s}\,\fsd\fs
\label{eq:Heff_at_f}\\
H_\theta^{at} & = & \frac{U}{2}\,\hL^2 +h\,\left(\hL-\frac{N}{2}\right)
\label{eq:Heff_at_theta}
\end{eqnarray}
The Lagrange multiplier $h$ is determined by the average constraint equation:
\begin{equation}
\langle\hL\rangle_{h} = N\left[n_F(\eps-h)-\frac{1}{2}\right]
\label{eq:atomic_h}
\end{equation}
in which $\langle\hL\rangle_{h}$ is the average of $\hL$ in the
Hamiltonian (\ref{eq:Heff_at_theta}):
\begin{equation}
\langle\hL\rangle_h = \frac{1}{Z_{\theta}}\,\sum_{\ell=-\infty}^{+\infty}
\,\ell\,e^{-\beta\,E_\ell}
\end{equation}
with: $E_\ell=U\ell^2/2+h\ell$ and $Z_{\theta}=\sum_\ell e^{-\beta E_\ell}$.
Solving (\ref{eq:atomic_h}) for $h$ as a function of $\eps$ and temperature
$T=1/\beta$ yields the dependence of the total charge within this approximation:
\begin{equation}
\langle\hQ\rangle = N\,n_F\left[\eps-h(\eps,T)\right]
\label{eq:approx_charge}
\end{equation}
We need to compare this approximation to the exact result (\ref{eq:exact_charge}) in the
atomic limit. A graphical representation (Fig.~\ref{fig:solve_constraint})
is useful in order to understand the solution of (\ref{eq:atomic_h}).
At $T=0$, one finds $h=\eps$, as long as $0<Q<N$. The exact
dependence of the average charge $Q$ upon $\eps$ at $T=0$ is correctly reproduced
by our approximation, corresponding to the ``Coulomb staircase'':
\begin{equation}
\label{eq:staircase}
Q = \frac{N}{2} + \ell \,\,\,\mbox{for}\,\,\,
-U\,\frac{2\ell+1}{2}<\eps<-U\,\frac{2\ell-1}{2}
\end{equation}
Note that $h-\eps$ vanishes linearly with temperature according to:
$h=\eps-T\ln(N/Q-1)+\cdots$: this is why the full Coulomb staircase can be
reproduced with a single Fermi factor in (\ref{eq:approx_charge}).
At finite temperature, our approximation does not coincide with the exact
result for $Q_{at}(\eps,T)$, but deviations are only sizeable for temperatures
comparable to $U$, which is not a severe limitation in practice.

\subsubsection{Spectral functions}

We now study the consequences of the approximate treatment of the constraint for
the Green's function and spectral function. Following (\ref{eq:Heff_at_f},\ref{eq:Heff_at_theta}),
the quantum rotor and auxiliary fermion degrees of freedom are described by two independent
Hamiltonians, so that the Green's function of
the physical electron $G_d(\tau)\equiv\,-\langle T\ds(\tau)\dsd(0)\rangle$
factorizes into:
\begin{equation}
G_d(\tau) = G_f(\tau)\,G_{\theta}(\tau)
\label{eq:Gd_tau}
\end{equation}
with $G_\theta(\tau)\equiv\langle\exp{i\left[\theta(0)-\theta(\tau)\right]}\rangle$.
Equivalently, the physical electron spectral function is given by:
\begin{equation}
\rho_d(\omega) = - \int d\omega'\, \rho_f(\omega')\rho_\theta(\omega-\omega')\,
\left[n_F(-\omega')+n_B(\omega-\omega')\right]
\label{eq:rhod}
\end{equation}
Let us consider $T=0$, and $\eps$ in the range corresponding to the plateau of charge
$Q$ in the Coulomb staircase. The ground-state energy is $E_Q=U(Q-N/2)^2+\eps Q$ and its
degeneracy is $\binom{N}{Q}$. The two excited states obtained by adding or removing a particle
correspond to transition energies:
$\Delta_{\pm}=E_{Q\pm1}-E_Q=\pm\eps\pm\,U(Q-N/2\pm 1/2)$. When acting with $\dsd$ on the
ground-state, only those ground-state components which do not already contain $\sigma$
contribute, and there are $\binom{N-1}{Q}=(1-\frac{Q}{N})\binom{N}{Q}$ such components.
Similarly, when acting with $\ds$, only the components in which $\sigma$ is occupied
contribute, and there are $\binom{N-1}{Q-1}=\frac{Q}{N}\binom{N}{Q}$ of them. From these considerations,
we see that the exact spectral function reads, at $T=0$:
\begin{equation}
\rho^{at}_d(\omega)\,=\,\left(1-\frac{Q}{N}\right)\,\delta(\omega-\Delta_{+})\,+\,
\frac{Q}{N}\,\delta(\omega+\Delta_{-})
\label{eq:rhod_at}
\end{equation}
These two atomic transitions are the precursors of the Hubbard bands in the solid. Note that
they have unequal weights, except at half-filling $Q=N/2$. At finite temperature, additional peaks
appear (except for $N=2$), corresponding to transition between two excited states (with exponentially
small weight for $T\ll U$).

Remarkably, the expressions (\ref{eq:Gd_tau},\ref{eq:rhod}) in which the quantum rotor and
auxiliary fermions are treated as decoupled, do reproduce this exact result at $T=0$.
The easiest way to see this is to notice that, at $T=0$,
$G_f(\tau)=-(1-Q/N)\theta(\tau)+(Q/N)\theta(-\tau)$, since $n_F(\eps-h)=Q/N$.
The rotor Green's function $G_\theta$ is $e^{-\Delta_{+}\tau}$ for $\tau>0$ and
$e^{\Delta_{-}\tau}$ for $\tau<0$. Substituting into (\ref{eq:Gd_tau}), this corresponds
to the exact expression (\ref{eq:rhod_at}). Alternatively, one can use the
expressions of the $T=0$ spectral functions into (\ref{eq:rhod}):
$\rho_f(\omega)=\delta(\omega-\eps+h)$ and
$\rho_\theta(\omega)=-\delta(\omega-\Delta_{+})+\delta(\omega-\Delta_{-})$,
keeping in mind that $n_F(\eps-h) = Q/N$ while
$n_B(\omega-\eps+h) = -\theta(-\omega)$ as $T\rightarrow 0$.
Again, deviations between the approximate
treatment and the exact results are found at finite temperature, but remain
small for $T\ll U$.
Let us emphasize that, because the rotor Green's function $G_\theta$ is a continuous function
at $\tau=0$, with $G_\theta(\tau=0)=1$, the factorized
approximation (\ref{eq:Gd_tau}) insures that the physical (d-electron) spectral function
is correctly normalized with total spectral weight equal to unity.

To summarize, we have found that treating the constraint on average reproduces
accurately the atomic limit at $T=0$, both regarding the Coulomb staircase dependence
of $Q$ vs. $\eps$, and regarding the spectral function. This is a key point
for the methods introduced in this article, which allows them to describe
reasonably the high energy features of strongly correlated systems.

\subsubsection{Functional integral formulation}

We briefly introduce here a functional integral formalism for the $\fsd$ and $\theta$
degrees of freedom, and derive the action associated with (\ref{eq:Hat}).
This is simply done by switching
from phase and angular-momentum {\it operators} $(\hat{\theta},\hat{L})$ to {\it fields}
$(\theta,\partial_\tau \theta)$ depending on imaginary time $\tau\in[0,\beta]$,
with $\theta(\beta)=\theta(0)+2\pi\,n$.
The action is constructed from $S \equiv \int_0^\beta \!\! d\tau [ - i \hL\,
\dt \theta + H + f^\dagger \dt f ]$, and an integration over $\hL$ is performed.
It is also necessary to introduce a
Lagrange multiplier $\widetilde{h}$ in order to implement the constraint $\hat{L} =
\sum_\s \fsd \fs - N/2$. We note that, because of the charge conservation on the
local impurity, $\widetilde{h}$ can be chosen to be independent of time, with
$\widetilde{h} \in [0,2\pi/\beta]$.
This leads to the following expression of the action:
\begin{equation}
\nonumber
S_{\mr{at}}  = \inte \sum_{\s} \fsd (\dt+\eps+i\widetilde{h}) \fs +
\frac{(\dt \theta + \widetilde{h})^2}{2U} -i \frac{N}{2} \widetilde{h}
\end{equation}
The constraint is implemented exactly provided $\widetilde{h}$
is integrated over.
The above approximation amounts to evaluate the integral by a saddle-point
approximation over $\widetilde{h}$,
and the saddle-point is found to be on the real axis, with $\widetilde{h}=ih$.

Finally, let us mention that, in a previous publication~\cite{SF_AG1}, we have explained
in detail the
connection between the rotor construction and the Hubbard-Stratonovich decoupling of
the interaction in the charge channel.

\section{The simplest mean-field approximation}
\label{sec:simplest}

In this section, we introduce a very simple mean-field approximation based on the
slave rotors variables. This approximation is similar in spirit to the condensation
of slave bosons in conventional slave boson mean-field theories. We illustrate this
approximation on two examples: the Anderson impurity model and the Hubbard model.

\subsection{Anderson impurity model}
\label{sec:anderson}

The Anderson impurity model describes a local orbital hybridized to a conduction
electron bath:
\begin{equation}
H = H_{at}
+\sum_{k\s} \ek c_{k\s}^\dagger c_{k\s}
+ V\,\sum_{k\s} (c_{k\s}^\dagger \ds  + \mr{h.c.})
\label{eq:AIM}
\end{equation}
This Hamiltonian can be rewritten in terms of the slave rotor and auxiliary
fermion variables:
\begin{eqnarray}
\nonumber
H & = & \frac{U}{2}\hL^2 + \eps\sum_\s \fsd\fs
+\sum_{k\s} \ek c_{k\s}^\dagger c_{k\s}\\
& + & V\,\sum_{k\s} (c_{k\s}^\dagger \fs\,e^{-i\theta}  + \mr{h.c.})
\label{eq:AIM_rot}
\end{eqnarray}
submitted to the constraint (\ref{eq:contrainte}). The simplest possible approximation
is to decouple the rotor and fermion variables, leading to two effective Hamiltonians:
\begin{eqnarray}\nonumber
H^{\mr{eff}}_f & = & (\eps-h)\sum_\s \fsd\fs
+\sum_{k\s} \ek c_{k\s}^\dagger c_{k\s}\\
& + & \veff\,\sum_{k\s} (c_{k\s}^\dagger \fs\,+ \mr{h.c.})
\label{eq:Heff_f_AIM}
\end{eqnarray}
\begin{equation}
\label{eq:Heff_rotor_AIM}
H^{\mr{eff}}_{\theta}= \frac{U}{2}\hL^2 +h\,\hL + K\,\cos\theta
\end{equation}
The parameters $\veff$, $K$ and $h$ in these expressions are determined by the coupled
self-consistent equations:
\begin{eqnarray}
\label{eq:Veff_AIM}
\veff & = & V \left< \cos\theta \right>_\theta \\
\label{eq:K_AIM}
K & = & V \sum_{k\s} \big< c_{k\s}^\dagger \fs +\fsd c_{k\s} \big>_f \\
\label{eq:h_AIM}
\langle\hL\rangle_\theta & = & N\left[n_F(\eps-h)-\frac{1}{2}\right]
\end{eqnarray}
in which the averages are calculated with the effective Hamiltonians above.

Let us first examine the particle-hole symmetric case $\eps=0$ ($Q=N/2$) in which
the solution of (\ref{eq:h_AIM}) is $h=0$.
The rotor sector is described by the effective Hamiltonian
(\ref{eq:Heff_rotor_AIM}) corresponding to the Schr\"{o}dinger equation:
\begin{equation}
\left[-\frac{U}{2} \frac{\partial^2}{\partial \theta^2}
+ K\;\cos\theta \right] \Psi(\theta) = E\,\Psi(\theta)
\label{eq:Hrotor}
\end{equation}
For $K=0$, the ground-state wave function is the state $|l=0\rangle$,
uniform on $[0,2\pi]$, corresponding to
maximal phase fluctuations and thus to the absence of charge fluctuations. This is
associated with the atomic limit, as explained above. As soon as the hybridization $V$ is
non-zero, we shall see that $K\neq 0$. The wave function is then maximum ($K<0$) for
$\theta=0\,,\,2\pi$, and $\langle \cos\theta\rangle$ acquires a non-zero expectation value.
This corresponds to a non-zero effective hybridization $\veff = V\langle \cos\theta\rangle$,
so that the auxiliary fermion effective Hamiltonian is that of a resonant level model.
This captures the physics of the Kondo effect, and the corresponding Kondo resonance at the
Fermi level.

Even though $V$ is a singular perturbation on the atomic limit, its effect can be
easily understood analytically in the present framework by treating the potential
energy $K\cos\theta$ perturbatively. To first order in $K$,
the ground-state wave function reads:
\begin{equation}
\big| \Psi_0^{(1)} \big> = \left|0\right> + \sum_{\ell \neq 0}
\frac{ \left<\ell\right| K \cos\theta \left|0\right>}
{E_{0} - E_\ell} \left|\ell\right>
\end{equation}
with $\langle\theta|\ell\rangle=e^{i\ell\theta}$ and $E_\ell=U\ell^2/2$. This yields:
\begin{eqnarray}
\nonumber
\left< \cos\theta \right>_\theta & = & \big< \Psi_0^{(1)}\big| \cos\theta
\big| \Psi_0^{(1)}\big> \\
& = & -2 K \sum_{\ell \neq 0} \frac{\left|
\left<0\right| \cos \theta \left|\ell\right> \right|^2 } { E_\ell} = - \frac{2K}{U}
\label{eq:cos}
\end{eqnarray}
Hence, using (\ref{eq:Veff_AIM}), one obtains $K=-U\veff/2V$, which yields the following
self-consistent equation for the effective hybridization $\veff$ when substituted into
(\ref{eq:K_AIM}):
\begin{equation}
\label{eq:veff1}
\veff\,=\,-\,\frac{2V^2}{U}\,\sum_{k\s} \big< c_{k\s}^\dagger \fs +\fsd c_{k\s} \big>_f
\end{equation}
The right-hand side of this equation is easily evaluated for the resonant level model
(\ref{eq:Heff_f_AIM}). For simplicity, we consider a flat conduction band
$\ek\in[-\Lambda,\Lambda]$, and focus on the universal regime:
$\Lambda \gg U \gg \Gamma$, with $\Gamma\equiv\pi\,V^2/2\Lambda$. To dominant order
in $1/\Lambda$, (\ref{eq:veff1}) reads:
\begin{eqnarray}\nonumber
1&=&- N \frac{2V^2}{U\Lambda}\,\int_{-\Lambda}^{0} \mr{d}\w \frac{\w}{\w^2 +
(\veff^2\pi/(2\Lambda))^2} \\
&=&N \frac{2V^2}{U\Lambda}\,\ln \left( \frac{2\Lambda^2}{\pi\veff^2}\right)
\end{eqnarray}
This yields the following expression for $\veff$ and for the width of the Kondo
resonance when $\Lambda \gg U \gg \Gamma$:
\begin{equation}
\label{eq:Gamma_eff}
\Gamma_\mr{eff} \equiv \frac{\pi \veff^2 }{2\Lambda} =
\Lambda \exp\left( - \frac{\pi U }{4N \Gamma} \right)
\end{equation}
This coincides with the exact expression~\cite{hewson}. The local orbital spectral function
obtained from (\ref{eq:Gd_tau}) reads:
\begin{equation}
\rho_d(\omega)\,=\, Z\,\frac{\geff/\pi}{\omega^2 + \geff^2}\,+\,\rho_d^{inc}(\omega)
\label{eq:rhod_AIM}
\end{equation}
The first term in this expression is the Kondo resonance, and carries a spectral weight
$Z=\geff/\Gamma=\left<\cos\theta\right>^2_\theta$.
It satisfies the Friedel sum-rule $\rho_d(\omega=0)=1/\pi\Gamma$.
Away from the particle-hole symmetric case ($\eps\neq 0$), the location of the resonance is
set by $\eps-h$, which is the renormalized impurity level familiar from conventional
slave-boson theories.
The rotor approximation does conserve total spectral weight, and therefore yields an
incoherent contribution to the spectral function with a weight $1-Z$. This incoherent
contribution is correctly centered around the atomic transitions, as explained above.
However, the width of these Hubbard bands is incorrectly described by the simple approximation
presented here, in which phase fluctuations are underestimated at short times. As a result,
the Hubbard bands have a bandwidth of order $\geff$ in this approximation (instead of the
expected, and much broader width, of order $\Gamma$). We note however that conventional
slave-boson approximations with a condensed boson neglect altogether the Hubbard bands at the
saddle-point level, and therefore the present approximation, simplified as it may be,
is preferable in this respect. An improved method for the treatment of phase degrees of freedom,
leading to a much more accurate description of the Hubbard bands, has been discussed
in previous publications~\cite{SF_AG1,florens_qdot_prb}. This method consists in a set of coupled
integral equations for the Green's functions of the auxiliary fermion and of the
slave rotor, in the spirit of the non-crossing approximation.

\subsection{Hubbard model}

\subsubsection{Slave rotor formulation}
\label{sec:slave_hubbard}

In this section, we consider the Hubbard model:
\begin{equation}
H\,=\,\sum_i H_{at}(i) - \sum_{\mr{ij},\s} t_{\mr{ij}}\, \disd \djs
\label{eq:Hubbard}
\end{equation}
which can be rewritten in terms of the rotor and auxiliary fermion variables as:
\begin{equation}
H\,=\,\sum_{\mr{i} \s} \eps \fisd \fis +
\frac{U}{2}\sum_{\mr{i}} \hat{L}_{\mr{i}}^2
- \sum_{\mr{ij}\s} t_{\mr{ij}}\, \fisd \fjs\,
e^{i(\theta_\mr{i}  - \theta_\mr{j})}
\label{eq:Hubbard_ftheta}
\end{equation}
Note that, in this context, $-\eps=\mu$ is the chemical potential controlling the
average density per site. Let us make a first approximation, which consists in decoupling the
rotor and fermion variables on links (besides treating the constraint on average, as
above), see~\cite{Kotliar_Liu} for a similar approach in the case of the t-J model.
We then obtain two effective Hamiltonians:
\begin{equation}
\label{eq:eff_fermi}
H_f = - \sum_{\mr{ij}\s} \hopeff\, \fisd \fjs\, + (\eps-h)\,\sum_{\mr{i} \s} \fisd \fis
\end{equation}
\begin{equation}
\label{eq:eff_XY}
H_{\theta} = - \sum_{ij} \Jeff \cos(\theta_i-\theta_j) +
\sum_{\mr{i}} \left( \frac{U}{2} \hat{L_i}^2 +h \hat{L}_i \right)
\end{equation}
corresponding respectively to free fermionic spinons with an effective hopping $\hopeff$ and
to a quantum XY-model for the phase variables with effective exchange
constants $\Jeff$. These effective parameters are determined by coupled self-consistent
equations:
\begin{equation}
\label{eq:eff_param}
\hopeff = \hop\,\langle \cos(\theta_i-\theta_j) \rangle_{\theta}\,\,\,,\,\,\,
\Jeff = \sum_\sigma \hop\, \langle\fisd\fjs\rangle_{f}
\end{equation}
in which the average values are calculated with the effective Hamiltonians above.
In addition, the Lagrange multiplier $h$ is determined from the constraint equation:
\begin{equation}
\langle \hL \rangle_{\theta} = \sum_\sigma
\left(\langle\fisd\fis\rangle_{f}-\frac{1}{2}\right)
\end{equation}
Let us emphasize that, in the decoupling leading to (\ref{eq:eff_fermi},\ref{eq:eff_XY}), we have
assumed that the average values $\langle\fisd\fjs\rangle_f$ and
$\langle\exp{i(\theta_i-\theta_j)}\rangle_\theta$ on a given bond are both real.
In fact, one could look for more general classes of solutions in which both
$\langle\fisd\fjs-\fjsd\fis\rangle_f\neq 0$
and $\langle\sin(\theta_i-\theta_j)\rangle_\theta\neq 0$. This would correspond to solutions
with orbital currents around a plaquette, as proposed by several
authors~\cite{Affleck_Marston}.
Spontaneous orbital currents are very naturally described using the slave rotor
method, but will not be considered further in this paper, which aims at
the general formalism.

\subsubsection{Simplest mean-field}
\label{sec:simplestMF}

In the next section, we shall investigate some physical consequences of
equations (\ref{eq:eff_fermi},\ref{eq:eff_XY},\ref{eq:eff_param}) which approximate
the Hubbard model by free spinons coupled self-consistently to an XY-model
for the phase degrees of freedom. We point out that the decoupling between
fermion and rotor degrees of freedom can be viewed as a controlled approximation
corresponding to a large-N limit of a multichannel model, as detailed in
appendix~\ref{sec:appendix}.

Here, in the same spirit as above,
we consider a further simplification, which consists in treating
the quantum XY model at the mean-field level.
In this framework, the phase degrees of freedom is described by
a mean-field Hamiltonian of independent sites:
\begin{equation}
H^{\mr{MF}}_{\theta} = \sum_{\mr{i}} \left[ \frac{U}{2} \hat{L_i}^2
+h \hat{L_i} + K \cos\theta_i \right]
\label{eq:H_hub_eff}
\end{equation}
with $K=-2\sum_j \Jeff \langle\cos\theta_j\rangle_\theta$.
Combining this with (\ref{eq:eff_param}) and calculating the average values
with the free-fermion Hamiltonian (\ref{eq:eff_fermi}),
we finally obtain the following self-consistency
equations for the variational parameters $K$ and $h$:
\begin{eqnarray}
\label{eq:K_hub}\nonumber
K & = & 2 N   \avcos\,\int \! \mr{d}\epsilon \, D(\epsilon)
\, \epsilon \; n_F\left(Z \epsilon + \eps - h\right) \\
\label{eq:L_hub}
\big< \hat{L} \big> & = &
N \int \! \mr{d}\epsilon \, D(\epsilon) \,
\left[n_F\left(Z \epsilon + \eps - h\right)-\frac{1}{2}\right]\\
\label{eq:Z}
Z & \equiv & \avcos^2_\theta
\end{eqnarray}
Finally, the relation between the chemical potential $-\eps$ and average
number of particle per site and color $n$ is given by:
\begin{equation}
\label{eq:number}
n\equiv \frac{1}{N}\sum_\sigma \langle\fsd\fs\rangle =
\int \! \mr{d}\epsilon \, D(\epsilon) \,
n_F\left(Z \epsilon + \eps - h\right)
\end{equation}
In these expressions,
$D(\epsilon)\equiv \int\frac{d^dk}{(2\pi)^d}\,\delta(\epsilon-\ek)$
is the density of states (d.o.s) of the band in the absence of interactions.
The auxiliary fermion (quasiparticle) Green's function reads:
\begin{equation}
\label{eq:greenf}
G_f(\vk,\iomn)^{-1} = \iomn -\eps+h - Z\, \ek
\end{equation}
We recognize $Z$ as the quasiparticle weight, which also determines
the quasiparticle mass enhancement $m^\star/m=1/Z$. These two quantities
are related
because of the simple single-site approximation made here.

At zero temperature, the number equation (\ref{eq:number}) implies that:
\begin{equation}
\label{eq:lutt}
h-\eps\,=\,Z\,\mu_0(n)
\end{equation}
in which $\mu_0$ is the chemical potential of the non-interacting system
such that $\int_{-\infty}^{\mu_0} d\epsilon D(\epsilon) = n$. From (\ref{eq:greenf}), it is
seen that the Fermi surface is located at $\ek=(h-\eps)/Z$, and thus (\ref{eq:lutt})
implies that Luttinger theorem is satisfied. In fact, within this simple approximation in
which the self-energy is independent of momentum, the Fermi surface is unchanged by interactions
altogether. The equations for $K$ and $h$, at $T=0$ and for a given density, simplify into:
\begin{eqnarray}
\big< \hat{L} \big> & = & N \left(n-\frac{1}{2}\right) \\
\label{eq:T0}
K & = & 2N \epsbar(n)\;\avcos_\theta
\end{eqnarray}
with $\epsbar(n)\equiv \int_{-\infty}^{\mu_0(n)} d\epsilon D(\epsilon)\epsilon$ the
average kinetic energy per electronic degree of freedom in the non-interacting model.

\subsubsection{Mott transition and orbital degeneracy}

\begin{figure}[htbp]
\begin{center}
\includegraphics[width=8cm]{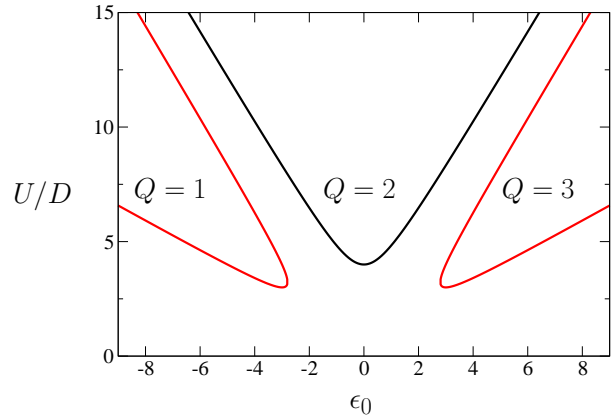}
\end{center}
\caption{Phase diagram for $N=4$ (two orbitals) at $T=0$, as a function of the
chemical potential $\eps=-\mu$ and the interaction strength $U/D$.
The three lobes correspond to the Mott insulator phases associated with half-filling ($Q=2$)
and quarter-filling ($Q=1,3$) respectively.}
\label{fig:phase_vs_eps}
\end{figure}

\begin{figure}[htbp]
\begin{center}
\includegraphics[width=6cm]{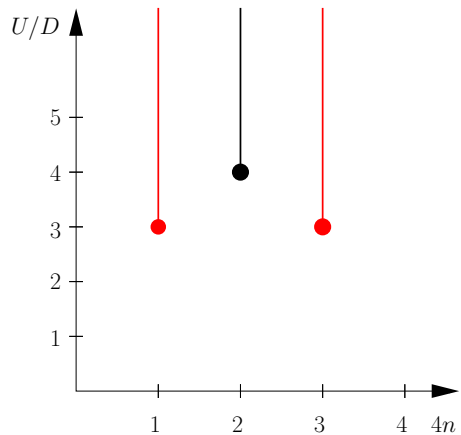}
\end{center}
\caption{Phase diagram in the ($n$,$U$) plane. The Mott insulator lobes
collapse to lines at commensurate fillings, when $U$ is larger than $U_c(n)$
(shown as dots).}
\label{fig:phase_vs_n}
\end{figure}

We expect a Mott transition to occur at each commensurate filling $n=Q/N$ ($Q$ being an
integer). This is associated with the vanishing of $Z$, and therefore the above
equations can be analyzed analytically close to the transition (where $Z$ is small)
from a perturbative analysis in $K$, similar to the one performed in
section~\ref{sec:anderson} for the Anderson model in the Kondo regime. The ground-state
wave function of $H^{\mr{MF}}_{\theta}$ in the insulating phase ($Z=K=0$) is
$e^{i\ell_n\theta}$ with $\ell_n=N(n-1/2)$. First-order perturbation
theory in $K$ yields:
\begin{eqnarray}
\nonumber
\left< \cos \theta \right>_\theta & = & 2 K \sum_{\ell \neq \ell_n}
\frac{\left| \left<\Psi_\ell \right| \cos \theta \left| \Psi_{\ell_n}\right> \right|^2 }
{E_{\ell_n} - E_\ell}\\
& = & - \frac{ 2 U K}{U^2 - 4 (U \ell_n + h)^2} + \mcal{O}(K^2)
\label{eq:cos_theta}
\end{eqnarray}
Since $Z$ vanishes at the transition, but $\mu_0(n)$ is finite, it follows from
(\ref{eq:lutt}) that $h=\eps$. For vanishing $Z$, the relation between $\eps$ and $n$ is
identical to that of the atomic limit, Eq.~(\ref{eq:staircase})
established in the previous section: $\ell_n=\mr{Int}\left[1/2-\eps/U\right]$ with
$n=1/2+\ell_n/N$. Finally, combining (\ref{eq:cos_theta}) and (\ref{eq:T0}), we obtain:
\begin{equation}
\label{eq:crit_boundary}
U_c(\eps)^2 - 4 \left[U_c(\eps)\ell_n+\eps\right]^2+4N\epsbar(n)U_c(\eps) = 0
\end{equation}
In this expression, $\ell_n$ and $n$ should be viewed as depending on the chemical
potential $\eps$ according to the relations just given. This expression determines the
boundary $U_c(\eps)$ between the metallic and Mott insulating phase in the $(\eps,U)$
plane. It is depicted for the case $N=4$ (two orbitals with spin) in Fig.~\ref{fig:phase_vs_eps}.
The condition $\partial\,U_c/\partial\eps=0$ determines the tip of each insulating lobe,
i.e the critical coupling $U_c(n)$ at which the insulating phase is entered as one increases
$U$ for a fixed commensurate density $n$. Differentiating (\ref{eq:crit_boundary}), it
is seen that this happens for $\eps=-Ul_n$, i.e precisely at the center of each step of the
Coulomb staircase. The critical coupling thus reads:
\begin{equation}
U_c(n)\, =\, 4N\,|\epsbar(n)|
\label{eq:uc_vs_n}
\end{equation}
The phase diagram in the $(n,U)$ plane is depicted in Fig.~\ref{fig:phase_vs_n} for
a flat d.o.s of half-width $D$, in which case $U_c=4NDn(1-n)$.
We see that the critical coupling is biggest at half-filling
$n=1/2$ ($Q=N/2$), which is expected since orbital
fluctuations are largest in this case. This conclusion may depend on the precise
shape of the d.o.s however (and in particular may not hold for densities of states such
that $D(-\epsilon)\neq\,D(\epsilon)$). The critical coupling increases linearly
with orbital degeneracy $N$. In fact, an analysis of the DMFT equations for large
orbital degeneracy was made in Ref.~\cite{florens_orbital_2002_prb},
and the exact behavior of the critical
coupling at leading order in $N$ found there is correctly reproduced by the simple
mean-field detailed here. It is also instructive to compare the present results
with that of the multi-orbital Gutzwiller approximation~\cite{lu},
which reads: $U_c^{\mr{GA}}=4(N+2)|\epsbar(n)|$. Our expression has the same behavior at large $N$,
but yields in general a smaller critical coupling: $U_c=U_c^{\mr{GA}}\,N/(N+2)$.
For small orbital degeneracies, we believe (on the basis of, e.g, DMFT results)
the Gutzwiller expression of $U_c$ to be quantitatively more accurate.

\begin{figure}[htbp]
\begin{center}
\includegraphics[width=8cm]{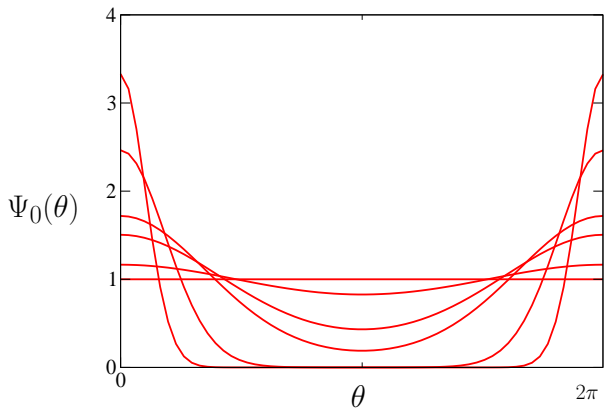}
\end{center}
\caption{Rotor ground state wave function $\Psi_0(\theta)$ with values of the
local interaction ranging from $U/U_c=0.01$ (peaked curve) to $U/U_c=1$ at the
Mott transition (flat curve).}
\label{fig:psi_hubbard}
\end{figure}

\begin{figure}[htbp]
\begin{center}
\includegraphics[width=8cm]{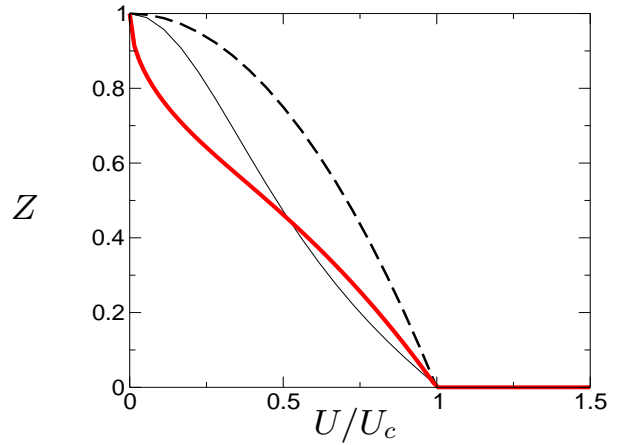}
\end{center}
\caption{Quasi-particle weight $Z$ as a function of $U/U_c$ at $T=0$; DMFT
calculation (thin line), rotor mean-field theory (thick line) and Gutzwiller
approximation (broken line).}
\label{fig:zcompare}
\end{figure}

The slave-rotor mean field equations are easily solved numerically by
determining iteratively the parameters $h$ and $K$. At each iteration, the
spectrum of the single-rotor Schr\"{o}dinger equations is computed (using
e.g a decomposition on the atomic basis states $e^{i\ell\theta}$).
In Fig.~\ref{fig:psi_hubbard}, the ground-state wave function $\Psi_0(\theta)$
is displayed for several values of $U$ at half-filling.
The curves nicely illustrate how one goes from
the insulator (in which case there are little charge fluctuations, and maximal
phase fluctuations so that the wave-function is delocalized over all $\theta$ values)
to the metal (in which case charge fluctuations become large at small $U$, and the
wave-function is peaked such as to limit phase fluctuations). The corresponding
quasi-particle weight is displayed in Fig.~\ref{fig:zcompare} as a function of $U/U_c$.
The simple slave-rotor mean-field is compared to the DMFT result and to the Gutzwiller
approximation (GA). It is seen that, close to the transition, the slave-rotor mean field
reproduces more accurately the DMFT answer than the GA. It is not very accurate
at weak-coupling however (even though $Z$ correctly goes to $Z=1$ at $U=0$, it has
an incorrect small-$U$ expansion). In fact, it is a quite general feature of this
slave-rotor mean field that the method is more accurate in strongly correlated regimes.

\begin{figure}[htbp]
\begin{center}
\includegraphics[width=7cm]{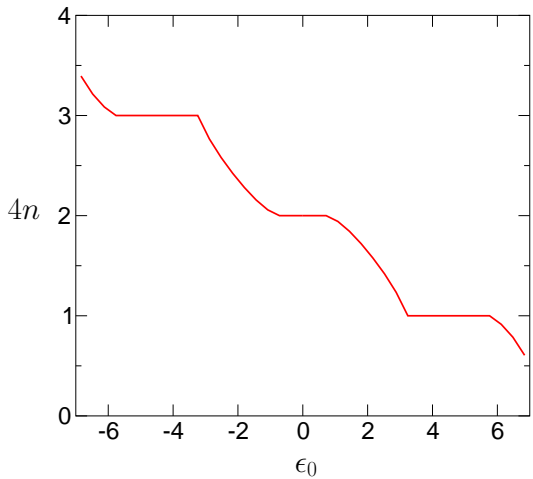}
\end{center}
\caption{Total occupancy $Q=4n$ as a function of $\epsilon_0$ for a value $U=4.5$ larger than
all critical interactions $U_c(n)$, in the two orbital case ($N=4$). The Mott
insulators are seen here as charge plateaus.}
\label{fig:n_vs_eps0}
\end{figure}

\begin{figure}[htbp]
\begin{center}
\includegraphics[width=8cm]{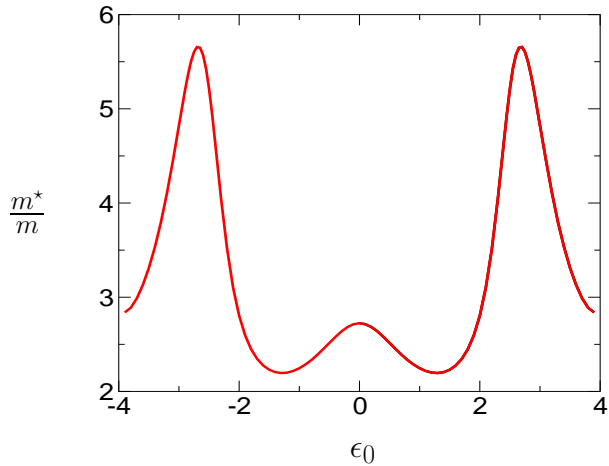}
\end{center}
\caption{Effective mass $m^\star/m$ for $U=2.5$ below all Mott transitions, in
the two orbital case ($N=4$).}
\label{fig:mstar_vs_eps0}
\end{figure}

In Fig.~\ref{fig:n_vs_eps0}, we plot the
number of particles as a function of the chemical potential
for $N=4$. The value of $U$ has been chosen to be bigger than the critical
couplings yielding an insulating state, for any commensurate filling. The curve
illustrates the plateaus found at each commensurate filling, the central one
(half-filling) being narrower (compare to Fig.~\ref{fig:phase_vs_eps}).
The effective mass enhancement ($=1/Z$) is also plotted in Fig.~\ref{fig:mstar_vs_eps0}
as a function of chemical potential for a smaller value of $U$, such that a metallic
phase is found at any filling. The curves illustrates how a largest
effective mass enhancement is found at low and high fillings $n=1/4,3/4$, and a comparatively
smaller close to  half-filling $n=1/2$ (again, this conclusion depends on the shape of the
d.o.s).

The description of the Mott transition obtained within this simplest
mean-field has many common features with the Brinkman-Rice (BR)~\cite{BR} one.
Indeed, the effective mass diverges at the transition and the quasi-particle
residue vanishes $(Z\sim 1-U/U_c)$ as in BR. There is one significant difference however,
which is that in the present description, the optical gap $\deltainf$ of the insulator {\it does not
coincide} with the chemical potential jump $\Delta\mu=-\Delta\eps=\mu(n+0^+)-\mu(n-0^+)$ for
infinitesimal doping away from
a commensurate filling. Indeed, within this simple mean-field, the spectral function of the
insulator is identical to that of the atomic limit (not surprisingly, the simple
mean-field with only two variational parameters describes the
charge fluctuations in the insulator in an oversimplified manner). As a result, the optical
gap simply reads
\begin{equation}
\label{eq:opt_gap}
\deltainf=U
\end{equation}
in our approach, and is therefore {\it not critical} at the
Mott transition. In contrast, the chemical potential jump vanishes continuously
at $U_c$. Indeed, solving (\ref{eq:crit_boundary}) for $\eps$ yields:
\begin{equation}
\label{eq:mu_jump}
\Delta\mu\,=\,U\,\sqrt{1-\frac{U_c}{U}}
\end{equation}
These features are very similar to those obtained within dynamical mean-field
theory~\cite{RMP_DMFT}.
This is not surprising, since the single-site mean field approximation to the XY-model
indeed becomes exact in the limit of infinite coordination of the lattice. Note however
that this is not the case of the approximation~(\ref{eq:eff_fermi})-(\ref{eq:eff_param})
which consists in decoupling the rotor and fermion variables (see section~\ref{sec:spatial}).
Within DMFT, the quasiparticle weight vanishes at a Brinkman-Rice like critical point $U_{c2}$ while
the optical gap of the insulator vanishes at a Hubbard-like critical point $U_{c1}$. As a result,
the strongly correlated metal close to the transition displays a clear separation of
energy scales: the quasiparticle coherence scale $\epsilon_F^\star\sim ZD$ being much smaller
than the (``preformed'') gap of the insulator $\deltainf$.
The simple mean-field of this section is in a sense a somewhat extreme simplification of this
picture, in which $U_{c2}=U_c$ and $U_{c1}$ is sent to $U_{c1}=0$ (this is consistent
with the known fact~\cite{florens_orbital_2002_prb} that $U_{c2}\propto N$ while $U_{c1}\propto\sqrt{N}$,
and that the simple mean-field becomes more accurate for large-$N$).

\section{Including spatial correlations and phase fluctuations}
\label{sec:spatial}

In this section, we go beyond the single-site mean-field approximation,
and investigate the physical consequences of the approximate description of the
Hubbard model introduced in Sec.~\ref{sec:slave_hubbard}. This description, summarized
by Eqs.(\ref{eq:eff_fermi}-\ref{eq:eff_param}), consists in a free
fermion model $H_f$ coupled self-consistently to a quantum XY-model
$H_{\theta}$ for the phase degrees of freedom.

\subsection{General considerations}
\label{sec:general}

Let us first emphasize some general aspects of this description, before turning
to explicit calculations.
The Hamiltonian for the phase degrees of freedom has two possible phases: a disordered
phase without long-range phase order, and a long-range ordered phase.
At zero-temperature, one expects a quantum phase transition from the ordered phase to the
disordered phase as the ratio $U/\Jeff$ is increased.
Since the Green's function of the physical electrons read, within this approximation:
\begin{equation}
G^d_{ij}(\tau-\tau') = G^f_{ij}(\tau-\tau')\,
\langle e^{i[\theta_i(\tau)-\theta_j(\tau')]}\rangle_\theta
\label{eq:Gd_hubbard}
\end{equation}
it is seen that the quasiparticle weight $Z$, associated with the limit of large-distance and large
time separation (low frequency), is given by:
\begin{equation}
Z\,=\,\langle\cos\theta_i\rangle^2_\theta
\end{equation}
Thus, the phase with long-range order for the rotors corresponds to the metal ($Z\neq 0$), while
the disordered phase correspond to the Mott insulator ($Z=0$).
Obviously, the description of the Mott metal-insulator transition that follows is closely analogous
to that of the superfluid-Mott insulator transition in the bosonic Hubbard model
\cite{fisher_bose,sachdev_book}.
Two remarks about this description of the metal and of the insulator are in order.
First, it is of course unphysical to think of a metal as having long-range phase coherence.
Naturally, this is only true of the saddle-point approximation in which the rotors and
spinon degrees of freedom are entirely decoupled. Fluctuations will induce interactions
between these degrees of freedom, restore inelastic scattering and thus destroy phase
coherence. The absence of inelastic scattering at the saddle-point level is
a well-known feature of slave-boson theories.
Note futhermore that despite the ordering of the rotors, the metallic phase
becomes a superconductor only when $\langle f^\dagger_{i\uparrow} f^\dagger_{j\downarrow}\rangle$
is also non-zero (i.e when there is spinon pairing).
Second, the insulator envisioned here is a {\it non-magnetic insulator} without any
spin or translational symmetry breaking, i.e a spin-liquid.
Even in the disordered phase, $\langle\cos(\theta_i-\theta_j)\rangle$ on a given
bond (e.g nearest-neighbour) is {\it non-zero} (it corresponds to the
energy density of the XY model). Therefore $\hopeff\neq 0$ in the insulating phase, so that
the spinons have a Fermi surface (with Luttinger volume). This also implies that
$\hopeff$ remains finite through the Mott transition and therefore that {\it the effective
mass does not diverge}, despite the fact that $Z\rightarrow 0$. These last remarks apply
to any finite dimension, but of course not to $d=\infty$. In this limit, the single-site
mean field of the previous section applies and $\langle\exp(\theta_i-\theta_j)\rangle =
\langle\cos\theta\rangle^2=Z$.
Finally, we emphasize that the non-magnetic nature of the insulator is
of course associated with the fact that the rotor degrees of freedom
are associated with the charge and are not appropriate to properly
describe spin ordering. Therefore, they are better suited to lattices
with strong frustration (or models with large orbital degeneracy) in
which a spin-liquid insulator is a realistic possibility.
Finally, because long-range order for the rotors corresponds to breaking a continuous
O(2) symmetry, a Goldstone mode will be present in the ordered (metallic) phase.
This mode is present in any finite dimension, but disappears in the $d=\infty$
limit. It corresponds to the zero-sound mode of the metal.
As we shall see, these long-wavelength modes play an important role: they change the
low-energy description of the transition as compared to the $d=\infty$ (DMFT) limit. As
a result, the separation of energy scales does not apply in a strict sense (the ``preformed''
gap found within DMFT is filled up with spectral weight coming from these low-energy modes).
As we shall see however, this spectral weight remains small in high dimensions (including
$d=3$), so that an approximate separation of scales still applies.

\subsection{Sigma-model representation: saddle-point equations in
the spherical limit}
\label{sec:spherical}

In order to perform explicit calculations with the quantum rotor
Hamiltonian (\ref{eq:eff_XY}), we shall use an approximation that
has proven successful in the context of quantum impurity models with
slave rotors~\cite{SF_AG1,florens_qdot_prb}. It consists in
replacing the quantum rotor $\exp(i\theta_i)$ by a complex bosonic field
$X_i(\tau)$ and to treat the constraint $|X_i|^2=1$ on average. Alternatively, this
can be viewed as extending the O(2) symmetry to O(M) and taking the large-M (spherical) limit.
This is a well known approximation to non-linear sigma models~\cite{sachdev_book}, which
preserves many qualitative features of the quantum phase transition. For details of the
formalism in the slave rotor context, see Ref.~\cite{SF_AG1}.
In the following, we focus on the half-filled case (since we are mainly interested in the
Mott transition), with a particle-hole symmetric d.o.s $D(\epsilon)$, so that
we can set $\eps=h=0$

The spinon and rotor (now X-field) Green's function
read~\footnote{A rescaling of the Coulomb repulsion
$U\rightarrow U/2$ was used in order to preserve the exact atomic limit, as
discussed in~\cite{SF_AG1}}:
\begin{eqnarray}\label{eq:Gf}
G_f(\vk,\iomn)^{-1}=\iomn-Q_f \ek \\ \label{eq:GX}
G_X(\vk,i\nu_n)^{-1}= \frac{\nu_n^2}{U} + \lambda + Q_X \ek
\end{eqnarray}
In these expressions, $\omega_n$ and $\nu_n$ are respectively,
fermionic and bosonic Matsubara frequencies, $\lambda$ is a Lagrange multiplier
associated with the constraint $\langle|X|^2\rangle=1$, while
$Q_f$ and $Q_X$ are the self-consistent parameters entering the
effective spinon hopping and XY coupling constants:
$Q_f=\langle\cos(\theta_i-\theta_j)\rangle=\langle X_iX^\star_j\rangle$,
$Q_X=\langle\sum_\sigma\fisd\fjs\rangle$.
The self-consistent equations which determine $\lambda$, $Q_f$ and
$Q_X$ read:
\begin{eqnarray}
\label{eq:gap}
1 &=& \inteps D(\epsilon)\;
\frac{1}{\beta} \sum_{n}
\frac{1} {\nu_n^2/U+\lambda+Q_X\epsilon}\\
\label{eq:Q}
D Q_f &=& -\inteps D(\epsilon)\; \epsilon
\frac{1}{\beta} \sum_{n}
\frac{1} {\nu_n^2/U+\lambda+\qbar\epsilon}\\
\label{eq:Qbar}
D \qbar &=& -2 \inteps D(\epsilon)\; \epsilon\;
n_F(Q_f\epsilon)
\end{eqnarray}
These expressions have been written here for a simple tight-binding band with
nearest-neighbor hopping $t_{ij}=t$ on a $d$-dimensional cubic lattice
($\ek=-2t\,\sum_{\alpha=1}^d\cos k_{\alpha}$). As above, $D(\epsilon)$ denotes
the band d.o.s, and $D=2d\,t$ is the half-bandwidth. For simplicity, we have set the
orbital degeneracy to $N=2$ in these equations.

\subsection{The Mott transition: Mott-Hubbard meets Brinkman-Rice}

In this section, we investigate the solution of these equations at zero-temperature.
This leads to a description of the finite-dimensional Mott transition that we
analyze in detail.

\subsubsection{The insulating phase}

Lets us note first that Eq.~(\ref{eq:Qbar}) readily determines
$\qbar$ at $T=0$:
\begin{equation}
D\qbar \;\under{T=0}\,=\,-2 \int_{-D}^{0} D(\epsilon)\,\epsilon\, d\epsilon \equiv 2 |\epsbar|
\end{equation}
From the form (\ref{eq:GX}) of the X-field Green's function, one sees that the
bosonic spectrum has a gap as long as
$\lambda-\qbar\,D = \lambda-2\abseps > 0$. In this case, there
is no long-range order for the
phase degree of freedom, and this corresponds to the insulating phase.
The insulating gap reads:
\begin{equation}
\gap = 2\,\sqrt{U(\lambda-\qbar D)}
\end{equation}
and we can rewrite Eqs.~(\ref{eq:gap},\ref{eq:Q}) as selfconsistent equations
for the gap $\gap$ and the renormalization of the spinon hopping $Q_f$. This
reads, at $T=0$:
\begin{eqnarray}
\label{eq:gapT0}
1 &=& \inteps D(\epsilon)
\frac{U}{\sqrt{\gap^2+8U\abseps(1+\epsilon/D)}} \\
Q_f &=& -\inteps D(\epsilon)\,
\frac{\epsilon}{D}\,\frac{U}{\sqrt{\gap^2+8U\abseps(1+\epsilon/D)}}
\label{eq:QT0}
\end{eqnarray}
These equations are valid in the insulating phase, when $\gap>0$. The gap vanishes
at a critical coupling $U_c$ obtained by setting $\Delta_g=0$ in (\ref{eq:gapT0}):
\begin{equation}
\label{eq:uc_dfini}
\frac{U_c}{\ucinf}\,=\,
\bigg[ \inteps \frac{ D(\epsilon)}{\sqrt{1+\epsilon/D}}  \bigg]^{-2}
\end{equation}
In this expression, $\ucinf=8\abseps$ is the critical coupling corresponding to the
$d=\infty$ limit, in agreement with expression (\ref{eq:uc_vs_n}) of the previous section
(with $N=2$). Note that, in the $d\rightarrow\infty$ limit, one must scale the hopping
as $t=t^\star/\sqrt{d}$, so that $D=2dt\propto\sqrt{d}\rightarrow\infty$ and
the r.h.s of (\ref{eq:uc_dfini}) goes to unity. The integral in (\ref{eq:uc_dfini}) is
smaller than unity in general, so that $U_c$ decreases as dimensionality is reduced.
We also note that in one dimension, this integral has a logarithmic singularity at
band edge, since $\pi\,D(\epsilon)=D/\sqrt{D^2-\epsilon^2}$, so that
Eq.~(\ref{eq:uc_dfini}) yields
$U_c^{d=1}=0$, which is indeed the exact result
for a half-filled Hubbard model with $N=2$~\cite{ref_1d}
(see however~\footnote{In the spherical X-field approximation, a finite
$U_c$ is not possible in $d=1$ because of the Mermin-Wagner theorem.
Had we kept the O(2) rotor variable, we would be able to correctly describe the
Berezinskii-Kosterlitz-Thouless nature of the Mott transition, even in the
approximation of decoupled spinons and rotors. A finite $U_c$ is possible in
$d=1$, as indeed found for larger orbital degeneracy,
see e.g the analysis of the case $N=4$ in~\cite{azaria_SU4}.}).
%

Substracting Eq.~(\ref{eq:gapT0}) from the same equation with $\Delta_g=0$
(which defines $U_c$), one obtains:
\begin{eqnarray}
 && \sqrt{\frac{\ucinf}{U_c}}-\sqrt{\frac{\ucinf}{U}} = \\
\nonumber
&& \inteps \, D(\epsilon) \left[\frac{1}{\sqrt{1+\epsilon/D}}
- \frac{1}{\sqrt{\gap^2/(U\ucinf)+1+\epsilon/D}}\right]
\end{eqnarray}
The expansion of this expression for small $\Delta_g$ depends on dimensionality.
For $d>3$, the integral $\int d\epsilon D(\epsilon) (1+\epsilon/D)^{-3/2}$ is convergent
at band edge $\epsilon=-D$, recalling that $D(\epsilon)\sim(D+\epsilon)^{d/2-1}$ near the
bottom of the band. In contrast, the small-$\Delta_g$ expansion is singular for $d<3$.
This analysis finally leads to the following behavior of the gap close to the critical point:
\begin{eqnarray}
\label{eq:gap_crit}\nonumber
\gap/U_c &\propto & \frac{U}{U_c}-1\,\,\mr{for}\,\,\, d>3\\
&\propto &\left(\frac{U}{U_c}-1\right)^{1/(d-1)}\,\,\mr{for}\,\,\, d<3
\end{eqnarray}
Hence we find that the exponent changes from its mean-field value $1/2$ for $d>3$ (as found
e.g in the single-site mean-field of the previous section and
the Gutzwiller approximation) to a non- mean field exponent for $1<d<3$.  Therefore,
$d=3$ corresponds to the upper critical dimension in this description of the Mott transition
(logarithmic corrections are found in that case).
Below $d=3$, the exponent $1/(d-1)$
corresponds to that of the large-$M$ limit of the quantum $O(M)$ model in $d$-dimensions,
i.e to that of the $d+1$-dimensional classical model.
Had we kept $O(2)$ quantum rotors,
we would have found $\Delta_g\sim (U/U_c-1)^{z\nu}$ with $z=1$ and
$\nu$ the correlation-length exponent of the $d+1$-dimensional classical XY model,
as in the case of the superfluid-insulator transition
of the Bose Hubbard model~\cite{fisher_bose}.

\subsubsection{The metallic phase}

For $U<U_c$, the gap closes and one enters the metallic phase. In this regime, the
constraint equation $\langle|X|^2\rangle=1$ can only be satisfied
by a {\it Bose condensation} of the X-field.
As in studies of quantum magnetism based on Schwinger bosons~\cite{arovas_auerbach},
Bose condensation in the spherical limit corresponds to the phase with long-range
order for the rotors.
In this phase, the constraint equation
(\ref{eq:gap}) has to be rewritten by isolating the $\vk=0$ mode in the Brillouin zone.
The Lagrange multiplier $\lambda$ sticks to the value $\lambda=\qbar D=2\abseps$ in this phase.
The full X-field Green's function thus reads at $T=0$:
\begin{equation}
\label{eq:GX_complet}
G_X(\vk,i\nu) =  Z \delta(\nu)\delta(\vk) +
\frac{1}{\nu^2/U+2\abseps(1+\ek/D)}
\end{equation}
The condensation amplitude $Z=\langle X\rangle^2$ is determined from the
constraint $\langle|X|^2\rangle =\sum_{\vk}G_X(\vk,\tau=0)=1$:
\begin{equation}
1 = Z + \inteps \,D(\epsilon)\; \sqrt{\frac{U}{8\epsbar(1+\epsilon/D)}}
\end{equation}
which simply reads, using (\ref{eq:uc_dfini}):
\begin{equation}
Z\,=\,1-\sqrt{\frac{U}{U_c}}
\end{equation}
This expression vanishes linearly, $Z \sim (U_c-U)/2U_c$, at the critical point
{\it for all dimensions} $d>1$.
%
The fact that there is no change of critical behavior for $Z$ at $d=3$, in contrast
to the gap, is due to the use of the spherical approximation. Had we kept O(2) rotors,
we would find $Z\sim (1-U/U_c)^{\nu(d-1+\eta)}$, with $\nu$ and $\eta$ the critical
exponents of the $d+1$-dimensional classical XY model. In the spherical approximation
$\eta=0$ and $\nu=1/(d-1)$ so that $Z\sim(1-U/U_c)$ also below $d=3$.

Hence, we have found that the quasiparticle weight and insulating gap {\it vanish at
a unique critical coupling}. As shown below, the gap $\Delta_g$ given by
(\ref{eq:gapT0}) is the gap in the single particle spectral density of the
insulator. It also coincides with the chemical potential jump
$\Delta\mu$ when the present approach is extended away from half-filling.
Hence, in this finite dimensional description of the Mott transition, we find a
unique critical point corresponding both to Brinkman-Rice physics~\cite{BR}
(vanishing of $Z$) and to Mott-Hubbard physics~\cite{Hubbard_MIT} (gap opening).
This is in strong contrast to the $d=\infty$ single-site mean field investigated in
the previous section, and to the DMFT picture~\cite{RMP_DMFT}. Below, we show that this
is due to long wavelength collective modes filling in the preformed gap, and investigate
in detail how the previous picture is recovered in the (singular) $d=\infty$ limit.

\begin{figure}[htbp]
\begin{center}
\includegraphics[width=8cm]{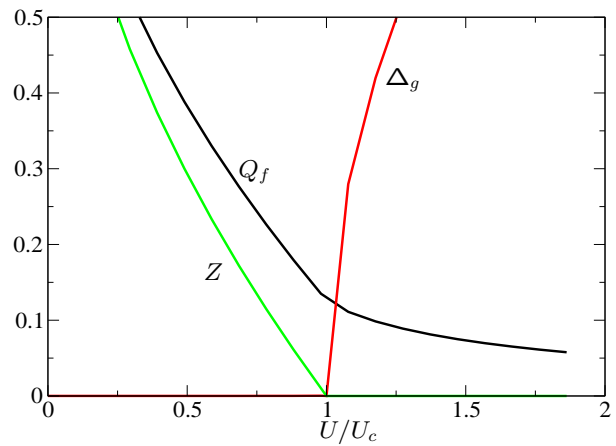}
\end{center}
\caption{Plot of the quasiparticle weight $Z$, the effective mass
renormalization $Q_f=m/m^{\star}$
and the Mott gap $\Delta_g$ as a function of $U/U_c$ across the
Mott transition in the three-dimensional case.}
\label{fig:scenario}
\end{figure}

\begin{figure}[htbp]
\vspace{0.5cm}
\begin{center}
\includegraphics[width=8.0cm]{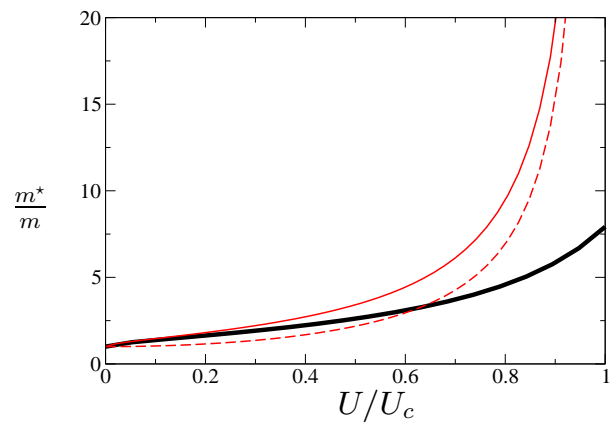}
\end{center}
\caption{Effective mass $m^\star/m=1/Q$ provided by the mean-field
equations~(\ref{eq:gap}-\ref{eq:Qbar}) for $d=3$ (bold line) and $d=\infty$
(thin line). For comparison, a DMFT-IPT calculation (dashed line) is also presented.}
\label{fig:Mstar}
\end{figure}

The equation (\ref{eq:Q}) for the renormalization $Q_f$ of the effective hopping must
be rewritten in the metallic phase to take into account the Bose condensed fraction.
At $T=0$, it reads:
\begin{equation}
\label{eq:mstar}
Q_f \,=\frac{m}{m^\star}\,= Z - \sqrt{\frac{U}{\ucinf}}\,\inteps D(\epsilon)
\frac{\epsilon/D}{\sqrt{1+\epsilon/D}}
\end{equation}
This expression makes very clear that the effective mass remains finite at the critical
point, while $Z\rightarrow 0$ (note that the integral in the r.h.s of (\ref{eq:mstar})
is negative so that $Q_f\geq Z$).  In the $d\rightarrow\infty$ limit, one recovers $m/m^{\star} = Z$, since
$D\propto\sqrt{d}\rightarrow~\infty$. This calculation can be extended to the
weakly doped Mott insulator at large $U$ and hole density $\delta$, with the result:
\begin{equation}
\frac{m^{\star}}{m} \sim \frac{1}{t/U+\delta} \sim \frac{1}{J/t+\delta}
\end{equation}
Hence, the present theory correctly captures
the magnetic exchange energy $J\propto t^2/U$, which quenches out the spin entropy (due to
spinon degrees of freedom) in the insulator and hence prevents the effective mass from
diverging at the Mott transition. This is expected from the fact that the spinons form a
dispersive band in the insulating state and thus have an entropy depending linearly
on $T$ at low temperature. These findings are entirely similar to the slave bosons mean-field
theories of the $t-J$ model~\cite{tJ_modes}. Fig.~\ref{fig:scenario} and
Fig.~\ref{fig:Mstar} illustrate graphically the physical quantities characterizing the Mott
transition which we discussed previously.

\subsubsection{Spectral functions and collective modes:
what are Hubbard bands made of.}

\begin{figure}[htbp]
\begin{center}
\includegraphics[width=8.0cm]{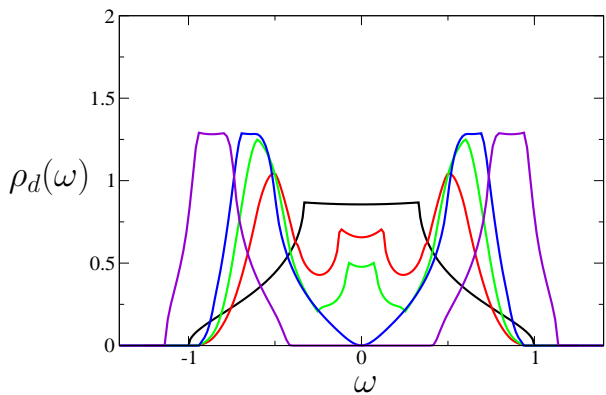}
\end{center}
\caption{Zero temperature local density of states across the Mott transition
for a three-dimensional cubic lattice, with $U=0$, $U_c/2$, $U_c$, $3Uc/2$
(for $D=1$).}
\label{fig:mott}
\end{figure}

The Green's function of the physical electron in the approximation of decoupled
spinons and rotors is given by (\ref{eq:Gd_hubbard}) as:
$G^d_{ij}(\tau)=G^f_{ij}(\tau)G^X_{ij}(\tau)$. Using (\ref{eq:GX_complet}) for the
X-field Green's function, this leads to:
\begin{equation}
\label{eq:Gd_metal}
G_d(\vk,i\omega) = \frac{Z}{i\omega - Q_f\,\ek} +
G_{d}^\mr{inc.}(\vk,i\omega)
\end{equation}
This expression is valid in the metallic phase. In Fig.~\ref{fig:mott}, we
display the $\vk$-integrated (local) spectral function
$\sum_{\vk}\rho_d(\vk,\omega)$,
as the Mott transition is approached, in the three-dimensional case.
The first term in (\ref{eq:Gd_metal}) corresponds to the coherent
quasi-particle. When summed over $\vk$, the quasiparticle contribution to the
local spectral function yields a peak $Z/Q_f D(\omega/Q_f)$. The spectral weight of this
peak is $Z$, its width is of order $Q_f D$ and its height is $ZD(0)/Q_f$. Hence,
its height goes to zero at the transition, while its width is reduced but remains finite
(Fig.~\ref{fig:mott}).
In the $d=\infty$ limit (where $Z=Q_f$) the zero-frequency density of states is
pinned at its non-interacting value, as known from the Brinkman-Rice picture and
the dynamical mean-field theory: {\it only in infinite dimension does the quasiparticle peak
disappear by narrowing down instead of collapsing}.

The incoherent contribution $G_d^{inc}$ comes from the convolution
of the free spinon Green's function
with the non-condensed contribution to the rotor Green's function,
$G_X^{inc}=1/[\nu^2/U+2\abseps(1+\ek/D)]$, which is the second term in (\ref{eq:GX_complet}).
The latter corresponds to bosonic collective modes dispersing according to:
\begin{equation}
\label{eq:disp_X}
\omega_X(\vk) = \pm\frac{\sqrt{U\ucinf}}{2}\sqrt{1+\frac{\ek}{D}}
\end{equation}
The incoherent contribution corresponds to the Hubbard band, which are well developed in
the correlated metal, as also predicted by DMFT. Note that the dispersing branch of bosonic excitations
is centered around $\pm\sqrt{U\ucinf}/2$.
Hence a measure of the typical energy scale associated with the distance
between the two Hubbard bands is:
\begin{equation}
\label{eq:preformed}
\Delta\,=\,\sqrt{U\ucinf}
\end{equation}
However, a key point is that this branch of collective modes extends to arbitrary
low frequency where it becomes the Goldstone mode of the broken symmetry. In the small
momentum limit, the dispersion relation (\ref{eq:disp_X}) reads:
\begin{equation}
\label{eq:sound}
\omega_X(\vk) \sim \sqrt{\frac{U\ucinf}{8d}}\,\,|\vk| \,\,\,(\vk\rightarrow 0)
\end{equation}
The corresponding density of states behaves as $\rho_X(\omega)\sim\omega^{d-1}$.
These long wavelength excitations are responsible for {\it tails of the Hubbard bands},
extending down to low- frequency.  This low-energy spectral weight due to collective modes is the origin of
the continuous closure of the Mott gap at the Brinkman-Rice transition.
In other words, the Hubbard bands are made of two kinds of contributions. The main part of
their spectral weight is associated with bosonic modes whose momentum is not small, so that $\omega_X$
is finite (of the order of $\Delta$).
In addition, the small weight in the ($\sim \omega^{d-1}$) tails at low-energy is associated with the
$\vk\simeq 0$ collective modes.
In the Mott insulator, all the bosonic modes are gapped, but the bottom of the bosonic
density of states has $(\omega-\Delta_g)^{d-1}$ tails which contribute to the tails of the
Hubbard bands, filling in the energy range between $\Delta_g$ and $\Delta$.

It is natural to interpret the bosonic collective modes as the zero-sound mode of the metal.
Indeed, these modes have been discussed previously by Castellani {\it et
al.}~\cite{tJ_modes}, in their study of fluctuations around the saddle-point of
conventional slave boson approaches.
These authors pointed out that the Mott transition is associated with the softening
of this collective mode, as also found here. In the present approach,
the collective modes appear on the same footing than the quasiparticles.

\subsubsection{The $d=\infty$ limit, separation of energy scales and ``preformed'' Mott gap}

It is instructive to understand more precisely what happens as the dimensionality is increased.
As clear from the previous discussion, the $d=\infty$ limit is singular in at least this
respect that the long-wavelength collective modes are absent.
Indeed the sound velocity in (\ref{eq:sound}) vanishes in this limit. In fact the bosonic modes
no longer have a dispersion: $\vk$-dependence disappears from
the dispersion relation (\ref{eq:disp_X}) since $D$ must be scaled as $D\propto\sqrt{d}$.
The bosonic spectral functions thus has two poles on top of the condensed fraction, which
leads, after performing the convolutions, to the following simple form of the physical
$\vk$-summed local spectral density (using also that $Q_f=Z$ in this limit):
\begin{equation}
\label{eq:dinfty_dos}
\rho_d^{d=\infty}(\omega)=D\left(\frac{\omega}{Z}\right) +
\frac{1-Z}{2}\left[D\left(\frac{\omega-\Delta}{Z}\right)+
D\left(\frac{\omega+\Delta}{Z}\right)
\right]
\end{equation}
In this expression, $\Delta$ is given by (\ref{eq:preformed}) and corresponds to
the typical separation between the Hubbard bands. It is sometimes referred in the framework of
DMFT as the ``preformed gap'' in the metallic state (i.e the Hubbard bands are well separated
from the central quasiparticle peak). $\Delta$ does not vanish at the Brinkman-Rice point and
beyond this coupling the insulator sets in with a finite gap $\Delta$. Note that, in the present
approximation where spinons and rotors have been decoupled, one simply has $\Delta=U$ in the insulator
(as found also in (\ref{eq:opt_gap})) and that, accordingly, the Hubbard bands in
(\ref{eq:dinfty_dos}) have vanishing width close to the transition. This
pathological result can be improved by including dynamical fluctuations of the
auxiliary particles, as shown in~\cite{SF_AG1}. Despite these oversimplifications,
the present approach does retain the main qualitative feature of DMFT, namely the separation of
energy scales at the Mott transition.

\begin{figure}[htbp]
\begin{center}
\includegraphics[width=8.0cm]{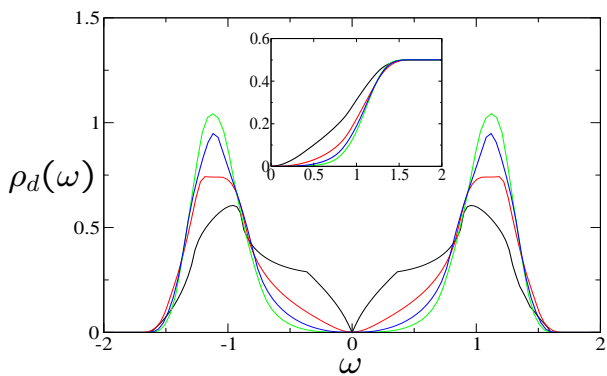}
\end{center}
\caption{Local spectral density at the critical point $U=U_c$,
for increasing dimensionality (top to
bottom curves at small $\w$: $d=2,3,4,5$).
The spectral weight associated with the low-frequency tails of the
Hubbard bands is seen to decrease as dimensionality increases. Correspondingly, the
separation of energy scales and the pre-formed gap become more and more apparent.
The inset shows the
integrated density of states $\int_0^\w \mr{d}\epsilon \, \rho_d(\epsilon)$
demonstrating this approximate separation of scales for $d \geq 3$.
(Note that the progressive narrowing of the main lobe of the Hubbard band as
$d$ increases is an artefact of the approximation in which spinons and rotors are
decoupled.}
\label{fig:rho}
\end{figure}

In Fig.~\ref{fig:rho}, we show how the large-d limit is {\it approached} by plotting the
local spectral density right at the critical coupling $U=U_c$, for increasing dimensionality.
This plot clearly reveals the two-components building up the Hubbard bands, with
the main part of the spectral weight centered around the ``preformed'' gap $\Delta$ and tails
extending down to low-frequency (down to the true gap $\Delta_g$ in the insulator), associated
with the long-wavelength collective mode. The inset demonstrates that as dimensionality
increases, the spectral weight in the tails becomes smaller, so that an approximate separation of
energy scales holds (and in fact already holds to a good approximation in $d=3$, while it
is no longer meaningful in two dimensions).

\section{Conclusion and perspectives}

In this paper, we have used the slave-rotor representation in order to construct
approximation schemes for strongly correlated fermion models. A theory of the
Hubbard model involving free fermionic spinons self-consistently coupled to a
quantum XY model has been developed. The quantum phase transition of the latter
corresponds to the Mott transition between a strongly correlated metal
and a Mott insulating spin-liquid with a spinon Fermi surface.
Both the gap in the spectral function and the quasiparticle weight vanish at the
critical point, while the effective mass remains finite.
In this picture, long-wavelength collective modes of the phase variable
play an important role.
They are responsible for low-energy tails of the Hubbard bands
(in addition to the main component of these bands, which are atomic-like
short-distance excitations). This has potential implications for
spectroscopic and tunneling experiments.
In infinite dimensions, these collective modes are suppressed and this limit
appears singular in this respect. Only in this limit is a strict separation
of energy scales recovered, as in the DMFT picture.

This raises some open questions associated with the physics of these
collective modes, which are physically associated with zero-sound. 
At the saddle-point level, where spinons and rotors do
not interact, these modes are undamped. The metallic state is described as
a perfect metal with no inelastic scattering. Interactions between rotors
and fermionic spinons will induce a Landau damping of these modes, and the metallic
state will loose phase coherence. One possibility is that this damping is large,
which would presumably weaken its effect and might restore some separation of
energy scales as in the DMFT picture. This calls for further work on the nature
of collective modes in a metal close to the Mott transition. In particular, the
restauration of gauge symmetry, broken at the saddle-point level, will have a
strong impact on the non-gauge invariant propagators, as already known for
single impurity models~\cite{Kondo_coleman}. However, it is yet not entirely
clear whether this effect will affect strongly the physical sector at low energy. 
One additional issue is that in a real material the screening deteriorates as one gets 
closer to the Mott insulating state~\cite{Mott_coulomb}. The ``acoustic plasmon'' mode of
the Hubbard model with short-range interactions will be pushed to higher energy and
this may also weaken its relevance for low-energy physics.
Another issue which will arise when taking into account the
interactions between spinons and rotors is the description of the insulator as
a non-magnetic spin-liquid. Stabilizing such a spin-liquid state beyond
saddle-point level is presumably possible only on a very frustrated lattice.
Even in this case however, it has been suggested recently~\cite{capone}
that a superconducting phase can intercalate between
the metal and the insulator, due to the proliferation of short range
spin singlets, therefore superseding the zero temperature Mott
transition. Nevertheless, some of our results, such
as the finiteness of the effective mass, should remain valid above the
low temperature ordered regions.

Finally, we point out that the slave rotor representation explored here
is a useful technical tool that can be applied to strongly correlated systems
in a variety of contexts. Because a single collective variable is introduced,
(which has a direct physical interpretation in connection with the local charge),
using this representation is generally simpler than other finite-U slave-boson
schemes provided one deals with a symmetric interaction.
Applications to mesoscopic devices and quantum impurity
models have been presented elsewhere~\cite{SF_AG1,florens_qdot_prb}.
Other potential applications are the effect of long-range or time-dependent
interactions~\cite{SF_LdM_AG}, or the interplay of disorder and interactions.
Interacting boson models can also be expressed with slave rotors (for a recent
application of variational approximations to the XY model, in the context
of bosonic models see~\cite{garcia-ripoll_mott}).
Although mean-field approximations for interacting bosons can be formulated in a
simple manner due to the commuting nature
of the physical degrees of freedom~\cite{sachdev_book}, the slave-rotor
representation might prove useful in the context of interacting cold atoms in
order to deal, for example, with boson-fermion mixtures.

\appendix
\section{Large-N limits and mean-field approximations}
\label{sec:appendix}

We discuss here how the different mean field approximations presented in this paper
can be formulated in terms of large $N$ limits of generalized Hubbard models.
Let us introduce a ``multichannel'' version of the Hubbard model,
based on spin-carrying
fermions $\fsd$ ($\sigma=1,\cdots,N$) and channel-carrying
phases $\theta_\alpha$ ($\alpha=1\ldots K$):
\begin{equation}
H\,=\,\sum_{\mr{i} \s} \eps \fisd \fis +
\frac{U}{2}\sum_{\mr{i}\alpha} \hat{L}_{\mr{i}\alpha}^2
- \frac{1}{K}\,\sum_{\mr{ij}\s\alpha} t_{\mr{ij}} \, \fisd \fjs\,
e^{i(\theta_{\mr{i}\alpha}  - \theta_{\mr{j}\alpha})}
\end{equation}
Two Hubbard-Stratonovich fields conjugate to $\sum_{\sigma}\fisd \fjs$ and
$\sum_{\alpha}e^{i(\theta_{\mr{i}\alpha}  - \theta_{\mr{j}\alpha})}$ can
be introduced in order to decouple the last term.
When both $N$ and $K$ are large, with a fixed ratio $K/N$, a saddle-point
applies which leads to the decoupled effective
Hamiltonians~(\ref{eq:eff_fermi}-\ref{eq:eff_XY}). This corresponds to a factorization
the hopping term {\it on bonds}, as shown by the effective
parameters~(\ref{eq:eff_param}). A similar remark applies in the usual
context of slave-bosons for the $t-J$ model: the mean-field approximation
investigated e.g in~\cite{Kotliar_Liu} correspond to a multichannel limit of:
\begin{equation}
-\frac{1}{K} \sum_{ij\sigma\alpha} t_{ij} \fisd\fjs\,b_{i\alpha}b_{j\alpha}
\end{equation}
Because the quantum XY model on the lattice is not easily investigated analytically,
we have performed in section~\ref{sec:spherical} a O(2M)
generalization of the phase part that leads to further simplications, while
allowing to deal with the model in finite dimensions. This can also be
seen as a direct large $N$,M limit of a SU($N$)$\times$O(2M) Hubbard model
(see~\cite{SF_AG1} for a related approximation concerning the Anderson model):
\begin{equation}
H = \sum_{\mr{i}\s} \eps \fisd \fis +
\frac{U}{2M} \sum_{\mr{i}\alpha\beta} \big(\hat{L}_{\mr{i}}^{\alpha\beta}\big)^2
- \sum_{\mr{ij}\s\alpha} \frac{t_{\mr{ij}}}{M} \, \fisd \fjs\,
X^*_{\mr{i}\alpha} X_{\mr{j}\alpha}
\end{equation}
introducing a complex field with M colors $X_{\mr{i}\alpha}$ subjected to the
spherical constraint: $\sum_\alpha |X_{\mr{i}\alpha}|^2 = M$.
In the previous expression, $\hat{L}_{\mr{i}}^{\alpha\beta}$ denotes the O(2M) angular momentum
tensor associated with the $X_{\mr{i}\alpha}$ vector~\cite{sachdev_book}.

Finally, we note that the simplest single-site mean field of section~\ref{sec:simplestMF}
can also be seen as a large-N limit of a generalized Hubbard model
which reads (note the different indices position and the scaling
of the hopping term):
\begin{equation}
H\,=\,\sum_{\mr{i} \s} \eps \fisd \fis +
\frac{U}{2}\sum_{\mr{i}\alpha} \hat{L}_{\mr{i}\alpha}^2
- \frac{1}{K^2}\sum_{\mr{ij}\s\alpha\alpha'} t_{\mr{ij}} \, \fisd \fjs\,
e^{i(\theta_{\mr{i}\alpha}  - \theta_{\mr{j}\alpha'})}
\end{equation}
This gives a {\it on site} factorization of the phase variables. Alternatively,
this can be seen as a large connectivity of the bond mean field
approximation~(\ref{eq:eff_fermi}-\ref{eq:eff_XY}).

\end{document}